# Slippage effect on laser phase error amplification in seeded harmonic generation free-electron lasers

Chao Feng[1,2], Haixiao Deng[1], Guanglei Wang[1], Dong Wang[1], and Zhentang Zhao[1*]

1 Shanghai Institute of Applied Physics, Chinese Academy of Sciences, Shanghai 201800, China

2 University of Chinese Academy of Sciences, Beijing 100049, China

Dao Xiang

SLAC National Accelerator Laboratory, Menlo Park, California 94025, USA

Abstract:

Free-electron lasers (FELs) seeded with external lasers hold great promise for generating high power radiation with nearly transform-limited bandwidth in soft x-ray region. However, it has been pointed out that the initial seed laser noise will be amplified by the frequency up-conversion process, which may degrade the quality of the output radiation produced by a harmonic generation scheme. In this paper, theoretical and simulation studies for laser phase error amplification in seeded FEL schemes with slippage effect taken into account are presented. It is found that, the seed laser imperfection experienced by the electron beam can be significantly smoothed by the slippage effect in the modulator when the slippage length is comparable to the laser pulse length. This smoothing effect allows one to preserve the excellent temporal coherence of seeded FELs in presence of large laser phase errors. For ultra-short UV seed lasers with FWHM around 16 fs, the slippage length in a modulator with ~30 undulator periods is typically comparable to the laser width; for longer seed laser pulses with FWHM around 80 fs, the slippage length can be made comparable to the laser width using a modulator tuned at the sub-harmonic of the UV seed laser. Three-dimensional simulations have been carried out for a soft x-ray facility using seed laser pulses with large frequency chirp and the results show that the sensitivity of the bandwidth of the seeded FEL to the initial frequency chirp can be significantly reduced by a proper design of the modulator such that the slippage length is comparable to the laser width. Our studies show that the tolerance on laser phase error for generating nearly transform-limited soft x-ray pulses in seeded FELs is much looser than previously thought and fully coherent radiation at nanometer wavelength may be reached with current technologies.

## I. Introduction

The recent success of self-amplified spontaneous emission (SASE) based x-ray free-electron laser (FEL) facilities [1, 2] has paved the way for novel types of experiments in many scientific disciplines. While the radiation from a SASE FEL has excellent transverse coherence, it typically has rather limited temporal coherence. There are many techniques (e.g. x-ray resonant inelastic scattering, spectroscopic studies of correlated electron materials, etc.) that could benefit from improved temporal coherence. To meet these scientific needs, various FEL seeding schemes such as high-gain harmonic generation (HGHG) [3] and echo-enabled harmonic generation (EEHG) [4, 5], have been proposed and experimentally demonstrated [6-9].

In seeded harmonic generation FELs, typically an external coherent seed laser pulse is first used to interact with electrons in a short undulator, called modulator, to produce energy modulation in the electron beam. This energy modulation is then converted into density modulation by a small chicane, called the dispersion section (DS). Taking advantage of the fact that the density modulation shows Fourier components at high harmonics of the seed, intense radiation at shorter wavelengths can be generated. Ideally, it is anticipated that the output radiation in a seeded FEL should inherit the properties of the seed laser with its bandwidth close to the Fourier transform limit. However, there are several challenges in implementing seeding schemes at extremely high harmonics. In particular, the initial insignificant errors compared to the seed wavelength may be amplified by the harmonic up-conversion process and will become large relative to a much shorter wavelength. For example, the electron beam shot noise will be amplified quadratically, which may overwhelm the external seeding source [10]. More recently, attentions have been turned to errors from the imperfection of the seed laser. It has been pointed out that if there is a phase chirp in the seed pulse, the chirp in the electron micro-bunching turns out to be roughly multiplied by the harmonic number $n$ [11, 12]. As a result, generation of nearly transform-limited radiation at 1 nm wavelength from a commercial 800 nm Ti:sapphire seed laser requires that the extra time-bandwidth product contributed by the seed phase chirp should be no more than one in a million of the ideal seed pulse [11], which is well beyond the state-of-the-art laser technology.

The conclusions in Ref [11, 12] were drawn under the assumption that the phase of the beam energy modulation directly copies the phase of the seed laser after the modulator. The seed power and frequency variations as a function of time due to the slippage effect in the modulator were neglected. This assumption is reasonable in the case of using an ideal seed laser with infinite pulse length and flat spectral phase distribution. However, for a realistic seed laser pulse with finite duration, the slippage effect on the energy modulation should be considered, especially for the case when the laser pulse width is comparable to the slippage length in the modulator.

As the continuous progress in laser technology has made ultra-short and high intensity laser pulses available, many seeded FEL facilities now use short seed laser pulses. For instance, the seed pulse duration of the FERMI seeded FEL is around 150 fs (FWHM) [13], and the Shanghai deep ultraviolet FEL uses an ultra-short seed pulse of about 80 fs (FWHM) [14]. FELs seeded with ultra-short lasers have been studied in Refs [15, 16] with a focus on wavelength tenability. In this paper, a new model that considers the slippage effect in the modulator is developed to describe the energy modulation with a frequency chirped short seed laser pulse. It is found that the noise induced by the seed laser can be smoothed by the slippage effect, when the slippage length in the modulator is comparable to the pulse length of the seed laser. For ultra-short UV seed lasers with FWHM around 16 fs, this condition is generally met with a modulator with ~30 undulator periods. For longer seed laser pulses with FWHM around 80 fs, we propose using a modulator tuned at the sub-harmonic of the UV seed laser to boost the slippage length to a similar level as the laser width. Three-dimensional simulations have been carried out for a soft x-ray facility to illustrate how the sensitivity of the bandwidth of the seeded FEL to the initial frequency chirp can be significantly reduced by a proper design of the modulator such that the slippage length is comparable to the laser width. Our studies show that the tolerance on laser phase error for generating nearly transform-limited soft x-ray pulses in seeded FELs is much looser than that suggested in [11, 12] and fully coherent radiation at nanometer wavelength may be reached with current technologies.

# II. Energy modulation with slippage effect

Here we consider a planar undulator with a sinusoidal magnetic field in the vertical direction and a period length $\lambda_u$. The undulator magnetic field of the modulator is

$$\vec{B}_y = B_0 \sin(k_u z)\vec{y}, \qquad (1)$$

where $B_0$ is the undulator peak magnetic field and $k_u = 2\pi / \lambda_u$ is the wave number of the undulator. The orbit of a relativistic electron in such a field is approximately a sine wave, and the velocity of the electron is given by

$$\vec{\upsilon}(t) = \upsilon_z(t)\vec{z} - \frac{Kc}{\gamma}\cos(\omega_u t)\vec{x}, \qquad (2)$$

where $\upsilon_z$ is the electron velocity in $z$ direction, the undulator parameter is $K \approx 0.934\, B_0 \lambda_u$ with $B_0$ in Tesla and $\lambda_u$ in centimeter, $c$ is the speed of light, $\omega_u$ is the angular frequency of the orbit and $\gamma$ is the is the relativistic electron energy. The electron's transverse velocity induced by the undulator magnet is,

$$\upsilon_x(t) = -\frac{Kc}{\gamma}\cos(\omega_u t). \qquad (3)$$

Since the velocity of the electron is constant, $\upsilon = c\sqrt{1 - 1/\gamma^2}$, $\upsilon_z$ can be calculated by

$$\upsilon_z(t) = c\sqrt{1 - 1/\gamma^2 - K^2 \cos^2(\omega_u t)/\gamma^2} \approx c - \frac{1 + K^2/2}{2\gamma^2}c - \frac{K^2 c}{4\gamma^2}\cos(2\omega_u t), \qquad (4)$$

So the average electron longitudinal velocity is

$$\bar{\upsilon}_z \approx c\left(1 - \frac{1 + K^2/2}{2\gamma^2}\right), \qquad (5)$$

Assuming a seed laser pulse with a Gaussian power distribution of rms width $\sigma_s$, central wavelength at $\lambda_s$ and a linear frequency chirp $\alpha$. The electric field distribution along the electron beam can be represented as

$$E(s) = E_0 e^{-s^2/4\sigma_s^2} e^{i(k_s s + \alpha s^2 + \phi_0)}, \qquad (6)$$

where $s$ is the position along the electron bunch, $E_0$ is the peak electric field of the seed laser, $k_s = 2\pi / \lambda_s$ is the wave number of the seed laser and $\phi_0$ is the initial carrier envelope phase of the laser. According to the resonant condition, the radiation overtakes the electron beam by one radiation wavelength per undulator period, which is called the slippage effect. Considering the slippage effect, the electric field and frequency distribution of the seed laser will vary with time, and Eq. (6) should be re-written as

$$E(s,t) = E_0 e^{-[s-p(t)]^2/4\sigma_s^2} e^{i\{k_s[s-p(t)] + \alpha[s-p(t)]^2 + \phi_0\}}, \qquad (7)$$

where $p(t) = (c - \bar{\upsilon}_z)t$ is the relative position of the seed pulse with respect to the electron beam. In a planar undulator, the electron has transverse wiggling motion and the longitudinal "figure-eight" oscillation. Such a trajectory gives rise to energy exchange between the electron and the laser electric field. The energy change of the electron can be calculated by

$$\gamma(s,t) = \frac{e}{mc^2} E(s,t) \cdot \upsilon_x(t), \qquad (8)$$

where $e$ and $m$ is the charge and mass of the electron. Integrating Eq. (8) with respect to $t$, we arrive at an expression for the energy modulation along the electron bunch after passing through the modulator:

$$\gamma(s) = \int_0^{L_{mod}/c} \frac{e}{mc^2} E(s,t) \cdot \upsilon_x(t) dt, \qquad (9)$$

where $L_{mod}$ is the length of the modulator.

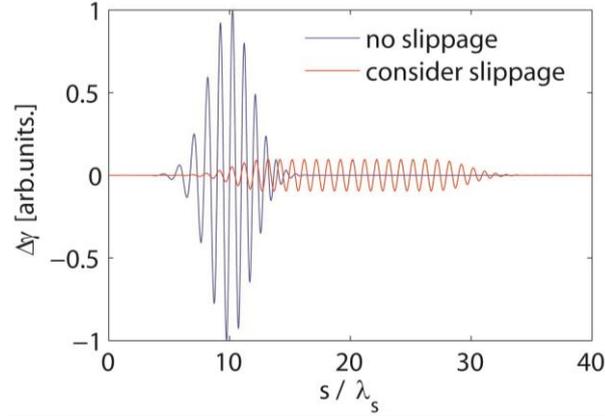

FIG.1. Energy modulation with an ultra-short seed laser pulse. The energy modulation amplitude is uniform in the central part with the slippage effect included using Eq. (7) (red line). For comparison, the energy modulation that directly copies the electric field is shown in blue line.

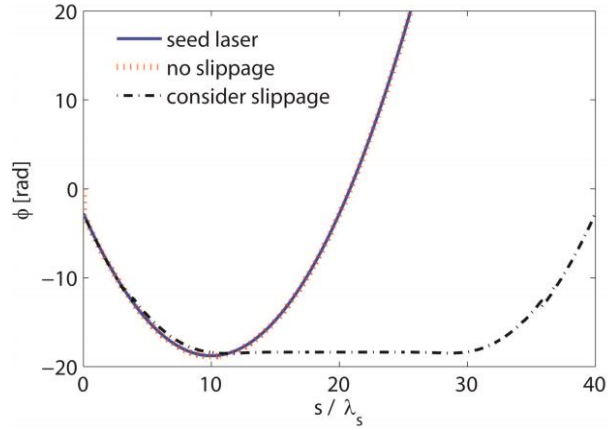

FIG.2. Spectral phase distributions of the seed laser and energy modulations: quadratic phase in the seed laser (blue line) results in quadratic phase in the energy modulation when neglecting the slippage effect using Eq .(6) (red dots); when considering slippage effect using Eq. (7) (black dash-dotted line) the phase in the central part of the energy modulation is nearly flat.

When using of an ultra-short seed pulse with the pulse length $\sigma_s$ comparable with the slippage length $N\lambda_s$ in the modulator, where $N$ is the period number of the modulator, part of the electron beam will be slipped over by the whole seed pulse and interacts with each cycle of the seed laser. So the energy modulation amplitude and the phase of the energy modulation will be averaged by the whole seed laser instead of directly copying the distribution of the seed pulse. Fig. 1 and Fig. 2 give the energy modulation results when using an ultra-short seed laser with FWHM pulse length of $3\lambda_s$ and a relatively large frequency chirp $\alpha = 0.16/\lambda_s^2$. The period number of the modulator is $N = 20$. For comparison of different models, we plug Eq. (6) and Eq. (7) into Eq. (9) to calculate energy modulation with and without slippage effect, respectively. The results are illustrated in Fig.1. The energy modulation amplitude is significantly reduced when considering the slippage effect, which means that the peak power of the seed laser and the strength of the DS should be properly enhanced to obtain sufficient spatial bunching in the electron beam [15, 16]. It is also found that the slippage effect creates a flat region in the central part of the energy modulation, where the energy modulation amplitude and phase distribution are uniform. Fig. 2 gives the phase distributions of the seed laser and energy modulations along the electron bunch for different cases. When the slippage effect is neglected, the energy modulation has quadratic phase that directly copies the spectral phase of the linear chirped seed laser. With the slippage effect taken into account, the spectral phase distribution is nearly flat in the central part of the energy modulation.

To fully characterize the energy modulation in both the time and frequency domains and compare with the seed laser pulse, we use the time-frequency distribution functions, which are appropriate tools to interpret the instantaneous carrier frequency. The Wigner distribution (WD) function [17, 18] has the simplest form among the usually used time-frequency distribution functions and has a good marginal property:

$$W(s,\omega) = \int \gamma(s-x/2)\gamma^*(s+x/2)e^{-i\omega x}dx, \qquad (10)$$

Where * denotes the complex conjugate and $\omega$ is the carrier frequency of the energy modulation. Fig. 3 shows the WDs of the seed laser and energy modulation for different period numbers of the modulator. Fig. 3(a) gives the WD of the seed laser pulse, where a considerable linear frequency chirp is clearly seen. When $N=1$, a linear chirp appears in the energy modulation as shown in Fig. 3(b). This chirp is significantly reduced in the central part of the energy modulation when $N=10$ (Fig. 3(c)), and the duration of the energy modulation with constant frequency begins to increase as the modulator length increases (Fig. 3(d)), which creates a platform region with quite flat spectral phase.

From Fig. 2(c & d), one can see that with the slippage length longer that or comparable to the laser pulse width, the initial seed laser frequency chirp only leads to chirp in the lateral parts of the energy modulation. Note, with the modulation amplitudes in the lateral parts much smaller than that of the central part, the bunching at high harmonics would be negligible at these lateral regions. Therefore, the frequency chirp in the lateral parts will not significantly affect the final radiation bandwidth.

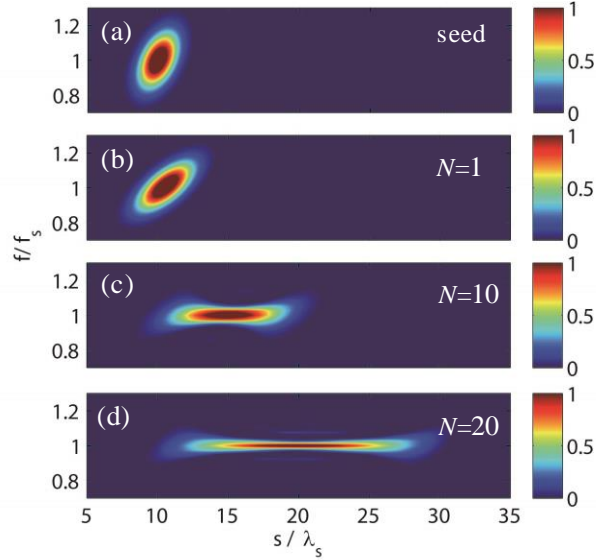

FIG.3. Wigner distributions of the seed laser pulse (a) and energy modulations (b-d) for different period numbers of the modulator ( $N=1, 10, 20$ ).

# III. HGHG and EEHG with short seed pulses

For HGHG scheme, after the modulator section, the electron beam is sent through a DS to convert the energy modulation into density modulation. The density modulation of the electron beam can be measured by the bunching factor [3], which has a maximum value of unity. The final electron density distribution after DS determines the properties of the radiation.

Consider an ideal beam with constant energy and current, assuming an initial Gaussian beam energy distribution with an average energy of $\gamma_0$ and rms energy spread $\sigma_\gamma$, the initial longitudinal phase space distribution can be written as

$$f_0(p) = \frac{N_0}{\sqrt{2\pi}} e^{-\frac{p^2}{2}}, \qquad (11)$$

where $N_0$ is the number of electrons per unit length and $p = (\gamma - \gamma_0)/\sigma_\gamma$ is the energy deviation of a particle normalized to the rms energy spread. After interacting with the seed laser pulse, the electron energy is changed to $p + A(s)$, where $A(s) = \gamma(s)/\sigma_\gamma$ is the energy modulation amplitude, and the distribution function becomes

$$f_1(p,s) = \frac{N_0}{\sqrt{2\pi}} \exp\{-[p - A(s)]^2 / 2\} \qquad (12)$$

Sending the electron beam through the DS with the normalized strength $B = k_s R_{56} \sigma_\gamma / \gamma_0$ converts the longitudinal coordinate to $s + Bp/k_s$, where $R_{56}$ is the dispersive strength, and makes the final distribution function

$$f_{HGHG}(p,s) = \frac{N_0}{\sqrt{2\pi}} \exp\{-[p - A(k_s s - Bp)]^2 / 2\} \qquad (13)$$

Integration of the formula over $p$ gives the beam density distribution along the electron beam. Then the local bunching factor can be written as

$$b(s,n) = \frac{1}{\sqrt{2\pi}} \int_{s-l/2}^{s+l/2} ds \int_{-\infty}^{\infty} \exp(-ink_s s) \exp\{-[p - A(k_s s - Bp)]^2 / 2\} dp \bigg/ l, \qquad (14)$$

where $n$ is the harmonic number and $l$ is the length of the chosen part of the electron beam.

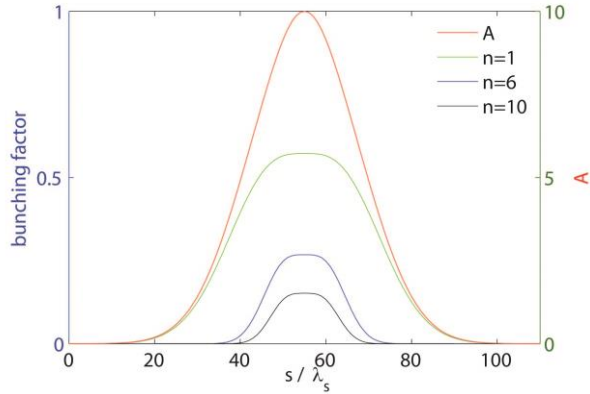

FIG.4. Energy modulation amplitude and corresponding local bunching factor distributions along the electron beam for different harmonics of the seed ($n$=1, 6 and 10).

The optimized strength of DS for $n$th harmonic bunching is $R_{56} \approx (1 + 0.81n^{-2/3})/A_m k_s$, where $A_m$ is the maximal value of energy modulation amplitude. Large bunching factor at high harmonics relies on the formation of sharp density spikes, which requires large energy modulation. For an HGHG FEL, the energy modulation amplitude $A$ should be $n$ times larger than the initial energy spread to give a considerable value of bunching factor at $n$th harmonic [19, 20]. However, as shown in Fig. 1, the energy modulation reduces for the lateral parts, which leads to broadened density spikes that cut off higher harmonics. This leads to increasingly short pulses as the harmonic number increases. Here we adopted an ultra-short seed laser with FWHM pulse length of $17\lambda_s$ and a frequency chirp $\alpha = 0.006 / \lambda_s^2$, the period number $N$ is chosen to be 30 to make the slippage length in the modulator comparable to the pulse length of the seed. We assume that the maximal energy modulation amplitude induced by the seed laser is 10 times larger than the initial energy spread. The local bunching factor distributions for different harmonics are calculated using Eq. (14), and the calculation results are shown in Fig. 4. The length of each slice is equal to the seed wavelength, $l = \lambda_s$. The bunching pulse length gets shorter at higher harmonics, which will change the harmonic spectral phase, as described in Ref [12]. While the bunching at the fundamental wavelength roughly follows the energy modulation amplitude and phase, the bunching at high harmonics only samples phase from the central part of the energy modulation. Fig. 5 shows the WDs of the seed laser, bunching factor distributions at fundamental and 10th harmonic of the seed. The fundamental bunching inherits the frequency distribution of the energy modulation and has frequency chirp at lateral parts, where the bunching factor is much smaller than the central part. For the 10th harmonic, only the energy modulation in the central part is large enough to generate sufficient bunching, and the phase of the bunching is quite flat in the central part. This slippage effect prevents the initial seed laser frequency chirp from broadening the final bandwidth of the high harmonic radiation.

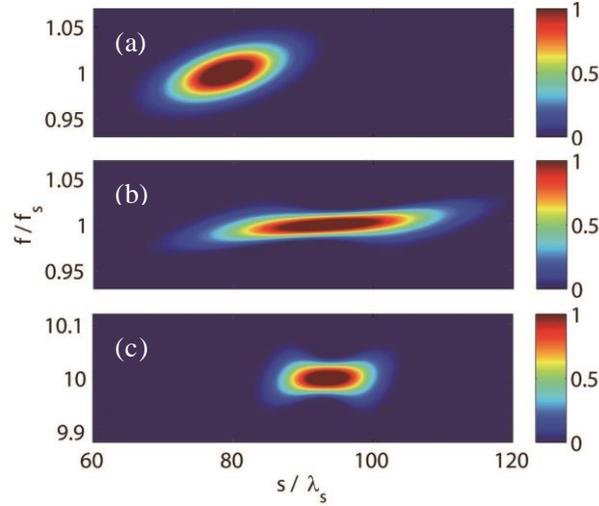

FIG.5. Wigner distributions of the seed laser (a), local bunching factor at fundamental (b) and 10th harmonic (c) of the seed.

Generally, the coherence property of the radiation could be quantified by the time-bandwidth product (TBP) factor, which can be simply defined as

$$T = \Delta k \Delta \tau, \qquad (15)$$

where $\Delta k$ is the spectral bandwidth and $\Delta \tau$ is the pulse duration. $T$ has minimal value $T_L$ for a transform-limited pulse and will grow as the phase error increases. Fig. 6 shows TBPs of the bunching factors as a function of the modulator period number for different harmonics of the seed. $T$ has been normalized by $T_L$. As can be seen from Fig. 6, the normalized TBP tends to decrease as the period number increases. The normalized TBP for fundamental and 6th harmonic bunching is close to one (transform-limited pulse) when $N$ is larger than 40. However, for 10th harmonic, the normalized TBP has a minimal value close to one when $N$=30. After that, the TBP start to grow due to the nonlinear effect of the frequency chirp in the seed laser, which can be calculated by Eq. (7) and Eq. (9). The phase distributions of the bunching in the central part of energy modulation for various harmonics are shown in Fig. 7. The period number is chosen to be 30 (Fig. 7(a)) and 50 (Fig. 7(b)) to perform a comparison. For $N$=30, the phase distributions are quite flat in the central part of energy modulation for fundamental and high harmonics, which will result in transform-limited pulses. For a larger period number, $N$=50, a small frequency chirp that is reverse to the chirp in the seed laser appears in the central part. This reversed frequency chirp is insignificant to the bunching at low harmonics, e.g. $n$<10, but will be amplified by harmonic number and significantly increase the TBPs for high harmonics. So it is necessary to properly set the period number of the modulator close to the optimized value for high harmonic generation. For our case, the optimized period number is around 30.

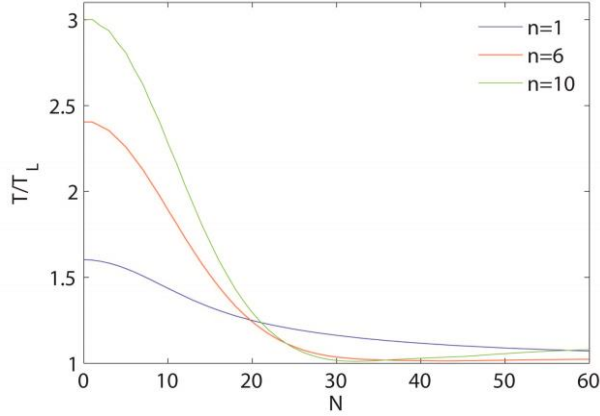

FIG.6. Normalized TBPs of the electron bunching as a function of the modulator period number for different harmonics of the seed ($n$=1, 6 and 10).

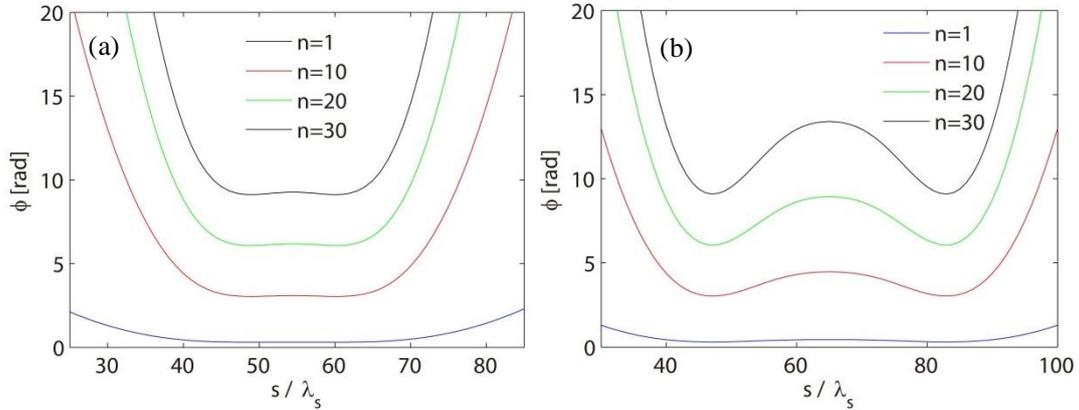

FIG.7. Phase distributions of the bunching in the central part of the energy modulation at various harmonics ($n$=1, 10, 20 and 30) for different period numbers of the modulator: (a) $N$=30; (b) $N$=50.

Fig. 8 gives the spectra of the electron bunching factors at the $10^{th}$ harmonic of the seed. The period number of the modulator is chosen to be $N$=30 to make the TBP of the $10^{th}$ harmonic bunching reach its minimal value. For convenience of comparison, four cases have been considered: flat phase and quadratic phase with and without slippage effect. It is found that the bandwidth of the spectral bunching for the quadratic phase case is about 3 times broader than that of the flat phase case when the slippage effect is neglected. However, the spectra are nearly the same for the flat phase case and quadratic phase case when the slippage effect is included, which implies that the frequency chirp in an ultra-short seed pulse may not significantly impact the electron bunching for HGHG with properly chosen period number of the modulator.

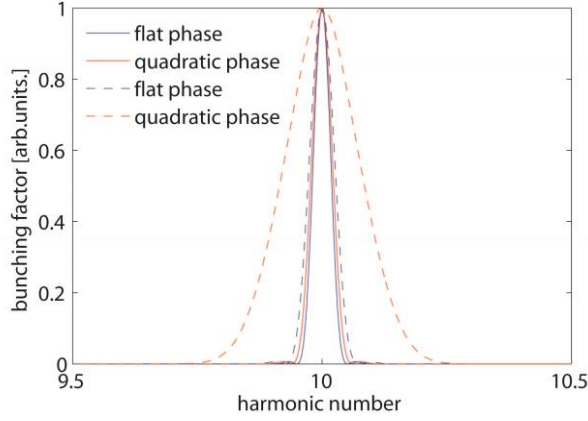

FIG.8. Bunching factor spectra for different cases: seed laser pulses with flat phase (blue solid line) and quadratic phase (red solid line) with slippage effect; seed laser pulses with flat phase (blue dashed line) and quadratic phase (red dashed line) without slippage effect.

The EEHG seeding mechanism shares many similarities with HGHG. In the standard EEHG scheme, the first laser-chicane combination filaments the electron beam in phase space. The second laser-chicane stage then simultaneously bunches each filament, resulting in multiple density spikes within each seed wavelength. Here we consider a scenario where the pulse length of the first seed laser is much longer than the second one. There are two benefits of this setup than using two ultra-short seed pulses: first, the long seeding pulse in the first modulator will provide a relativity uniform modulation and phase space distribution along the electron beam, which is useful to obtain a large bunching factor; second, this setup loosens the constraints on the timing control of these two seeding pulses.

The second stage of EEHG is similar to the HGHG process, with the energy separation of the filaments determining the final harmonic number. The frequency chirp in the seed laser will affect the two stages differently [12]: For the first stage, the chirp mainly distorts the separation of the filaments, which decrease the peak value of the bunching factor. However, because the second seed laser still phase locks each set of density spikes, the chirp from the first laser has little effect on the bandwidth of the bunching. Here we focus on the spectral phase of the bunching factor, so a flattop laser pulse in the first stage and a Gaussian pulse with rms length $\sigma_s$ in the second stage is assumed. Following the notation of Ref [5] and assuming the central frequencies of the two seed lasers equal, the final electron beam distribution function after the second DS can be written as

$$f_{EEHG}(p,s) = \frac{N_0}{\sqrt{2\pi}} \exp\{-[p - A_2(k_s s - B_2 p) - A_1 \sin[(k_s s - B_1 p - B_2 p) + B_1 A_2(k_s s - B_2 p)]]^2 / 2\} \quad (16)$$

where $A_2(s) = \gamma(s) / \sigma_\gamma$, and the local bunching factor distribution for EEHG is

$$b(n,s) = \frac{1}{\sqrt{2\pi}} \int_{s-l/2}^{s+l/2} ds \int_{-\infty}^{\infty} \exp(-ink_s s)\exp\{-[p - A_2(k_s s - B_2 p)$$
$$-A_1 \sin(k_s s - B_1 p - B_2 p) + B_1 A_2(k_s s - B_2 p)]^2 / 2\} dp / l \quad (17)$$

We assume that the energy modulation amplitudes for the two stages are $A_1 = 3$, $A_{2m} = 2$, where $A_{2m}$ is the maximal value of $A_2$, and the length of each slice is equal to the seed wavelength, $l = \lambda_s$. The properties of the second seed pulse are the same as that used in the HGHG case. The optimized strengths of the DSs for $n$th harmonic bunching factor can be calculated using the method given in Ref [5]. Fig. 9 shows the local bunching factor distribution along the electron bunch for different harmonic numbers. As the bunching factor is very sensitive to the energy modulation amplitude at high harmonics, the bunching distributions has sharp edges at lateral parts of the energy modulation. Fig .10 gives the Normalized TBPs of the electron bunching as a function of the modulator period number for different harmonics. It is found that the optimized period numbers for high harmonics are all around 30. For $n = 20$ and 30, the TPBs have peaks around $N = 50$ due to the reversed frequency chirp in the central part of energy modulation. After that the TPBs start to decrease again as the period number increase.

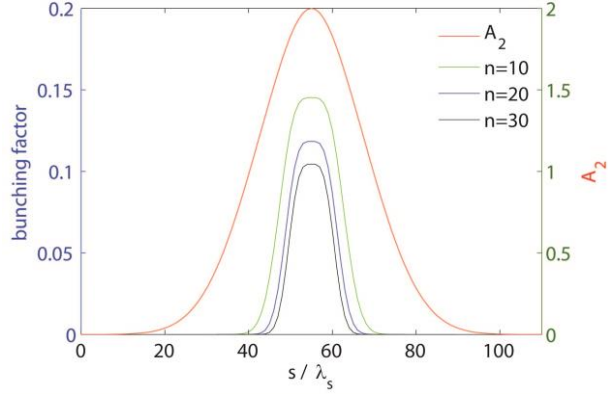

FIG.9. Local bunching factor distributions along the electron beam for different harmonics of the seed ($n$=10, 20 and 30).

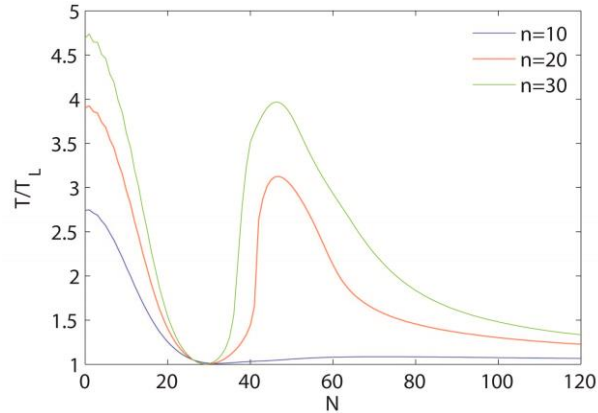

FIG.10. Normalized TBPs of the electron bunching as a function of the modulator period number for different harmonics of the seed ($n$=10, 20 and 30).

# IV. 3D simulations for SXFEL

To verify the theoretical results in previous sections, we have carried out 3D simulations using GENESIS [21] based on the nominal parameters of Shanghai Soft X-ray FEL (SXFEL) project [22]. The SXFEL test facility aims at generating 8.8 nm FEL from a 264 nm conventional seed laser through a two-stage cascaded HGHG or a single stage EEHG configuration. The electron beam energy of SXFEL is 840 MeV with emittance of 1mm-mrad and slice energy spread of about 84 keV. The beam peak current is 600 A.

For cascaded HGHG operation, the test facility converts the seeding laser at wavelength $\lambda_{seed} = 264nm$ to the FEL at 44nm with the first stage HGHG and it is followed by the second HGHG stage to produce the 8.8 nm soft x-ray. Here we only consider the first stage HGHG with harmonic up-conversion number of 6. We assume the pulse length of the seed laser is 16 fs (FWHM), and the period number of the modulator is 30. For comparison purpose, seed lasers with flat and quadratic spectral phase (linear chirp) distributions have been considered. For the quadratic phase case, we set $\alpha = 0.006 / \lambda_s^2$, which makes TBP of the seed pulse 1.6 times larger than that of a transform-limited pulse with the same pulse length. According to Ref [11, 12], the TBPs of the output radiations should be 2.4 and 4.7 times larger than that of transform-limited pulses for 6th and 30th harmonics when ignore the slippage effect.

The resonant wavelength of the radiator in the first stage is flexible, so we can make the radiator resonant at the fundamental and $6^{th}$ harmonic of the seed. The WDs of the fundamental (264 nm) and $6^{th}$ harmonic (44 nm) radiations at the very beginning of the radiator for the chirped case are shown in Fig. 11. As the FEL works in the coherent harmonic generation (CHG) regime [3], the radiation power is proportional to the square of the bunching factor. One can clearly see that the frequency chirps occur in the lateral parts of the radiation pulses, which agrees with the theoretical predictions as shown in Fig. 5. As the central part with large radiation power has flat phase distribution, the spectrum of the radiation for the quadratic phase case will have little difference from the flat phase case. Fig. 12 shows the FEL performance of the $6^{th}$ harmonic radiation for these two cases. It is found that the gain curves and spectra of the radiations are nearly the same. The bandwidths of the radiation pulses at saturation are both around 0.2%, which indicates that the initial frequency chirp in the seed laser does not lead to broadening of the bandwidth of the harmonic radiation of HGHG.

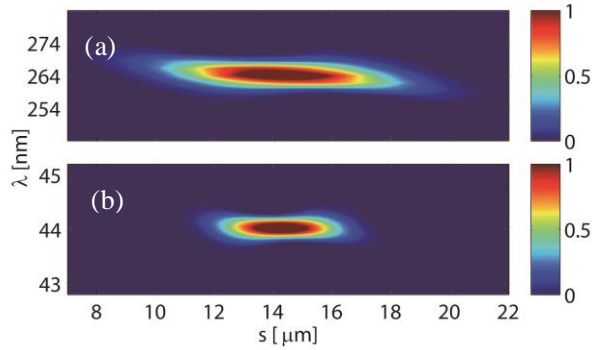

FIG.11. Wigner distributions of the coherent harmonic radiation at fundamental (a) and $6^{th}$ harmonic (b) of the seed.

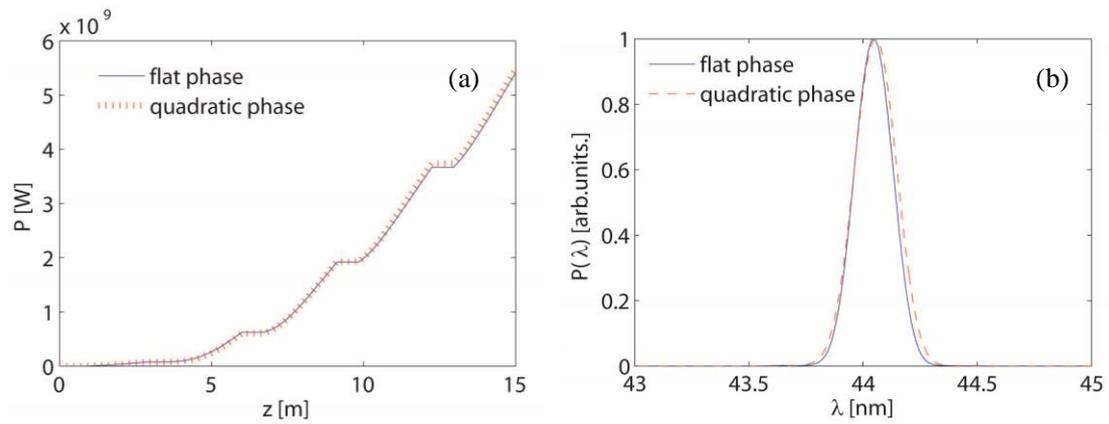

FIG.12. $6^{th}$ harmonic radiation performances of HGHG for the 16 fs seed laser pulses with flat phase (blue line) and quadratic phase (red dashed line): (a) FEL gain curves; (b) spectra at saturation.

For EEHG operation, we assume that the first seed laser pulse is longer than the electron bunch. The properties of the second seed laser are chosen to be the same as that used in HGHG simulations. The period numbers of the two modulators are both 30. The energy modulation amplitudes and dispersion strengths in our simulation are set to be $A_1 = 3.5, A_2 = 4, B_1 = 8.86, B_2 = 0.28$ to maximize the bunching factor at 30th harmonic of the seed. The corresponding seed laser peak powers are 40 MW and 80 MW, respectively. The performance of the radiation also has a very weak dependence on the chirp in the seed laser, as shown in Fig. 13. The bandwidths of the radiation pulses at saturation are also around 0.2%.

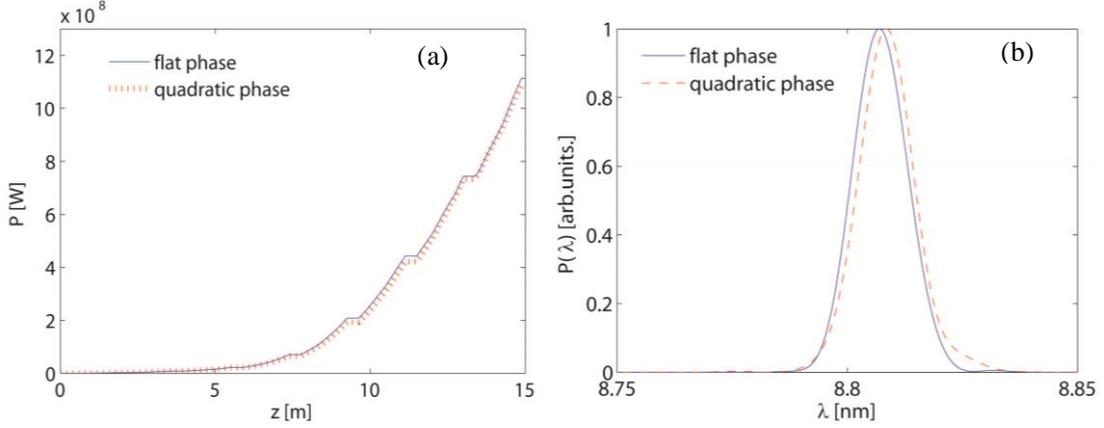

FIG.13. 30th harmonic radiation performances of EEHG for the 16 fs seed laser pulses with flat phase (blue line) and quadratic phase (red dashed line): (a) FEL gain curves; (b) spectra at saturation.

# V. Slippage-boosted by a sub-harmonic modulator

In the previous section, we have shown that nearly transform-limited high harmonic radiation pulses can be generated via seeded FELs, when the slippage length in the modulator is comparable to the pulse length of the seed laser. For a 16 fs seed laser pulse at 264 nm, the optimized period number is about 30, which is a reasonable value for modulators design. However, when using a longer seed laser pulse for the generation of narrower bandwidth radiation, this period number is not large enough to compensate the initial frequency chirp induced by the seed. Fig. 14 shows the simulated spectra of the 30th harmonic radiations of EEHG with 80 fs (FWHM) seed laser pulses. The frequency chirp in the seed laser pulse is $\alpha = 0.0006 / \lambda_s^2$ for the quadratic phase case, which makes TBP of the seed pulse 3 times larger than that of a transform-limited pulse with the same pulse length. The period number of the modulator is set to be 22 here. Other parameters are the same as that used in the previous section. It is found from Fig. 14 that the bandwidth of the 30th harmonic radiation for the quadratic phase case is about 4.5 times broader than that of the flat phase case. Although the TBP of the radiation pulse has already been reduced by the modulator with 22 periods, the period number still needs to be further increased by about 5 times to fully compensate the frequency chirp induced by the 80 fs seed laser pulse. However, when the period number is too large, e.g. larger than 50 for SXFEL, the modulator will no longer work in the small-gain regime. In this case the FEL interaction tends to wash out the fine structures in energy space and it will lead to a significant degradation to the quality of the electron bunch.

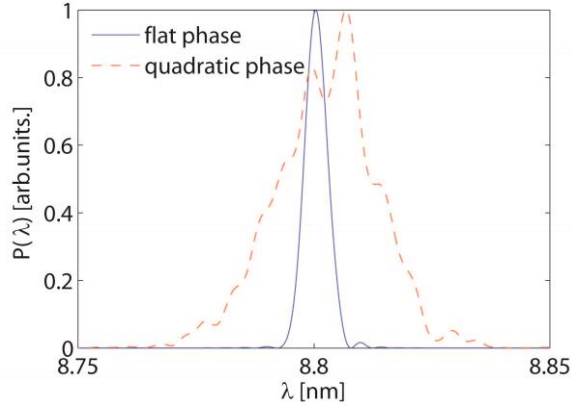

FIG.14. Spectra of 30[th] harmonic radiation pulses of EEHG for the 80 fs seed lasers with flat phase (blue solid line) and quadratic phase (red dashed line).

It has been proposed that undulator sections resonant at sub-harmonics of the FEL radiation can be used to increase the FEL slippage length, which will lead to a reduced bandwidth of a SASE-FEL [23]. Here, we apply the similar idea to the modulator of seeded FEL schemes for the slippage-boosting purpose. Instead of resonant at the fundamental of the seed, the modulator is tuned to an odd sub-harmonic: $\lambda_m = m\lambda_s$, $m = 3,5,7,...$, which will increase the slippage length by $m$ times in the modulator while simultaneously keeping the FEL interaction in the small-gain regime.

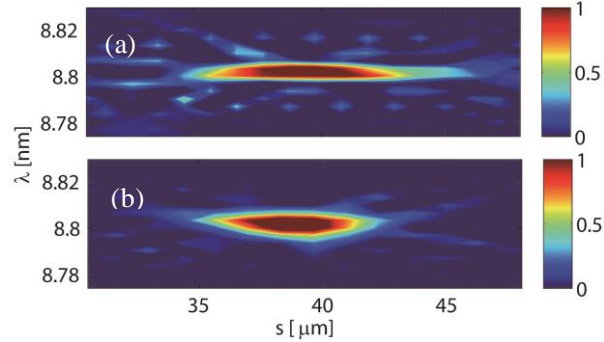

FIG.15. Wigner distributions of the 30[th] harmonic radiation at the beginning of the radiator of EEHG for the 80 fs seed laser with flat (a) and quadratic (b) phase distributions.

Here we make the second modulator of EEHG resonant at 1320 nm, which is 5 times longer than the seed wavelength. The peak power of the second seed laser is increased to 180 MW to generate the same energy modulation amplitude $A_2 = 4$. Fig. 15 shows the Wigner distributions of the 30[th] harmonic radiation at the very beginning of the radiator for the seed laser pulses with flat phase and quadratic phase. One can find that there are no chirps in the radiation for both these two cases, which implies that the initial chirp in the seed laser has been compensated by the sub-harmonic modulator. Fig. 16 gives the FEL performance for these two cases. The output peak powers for these two cases are at the same level. The bandwidths of the radiation pulses at saturation are nearly the same, both around 0.05%, which is about 4 times narrower than the output bandwidth of the normal EEHG with a 16 fs seed laser pulse (Fig. 13 (b)).

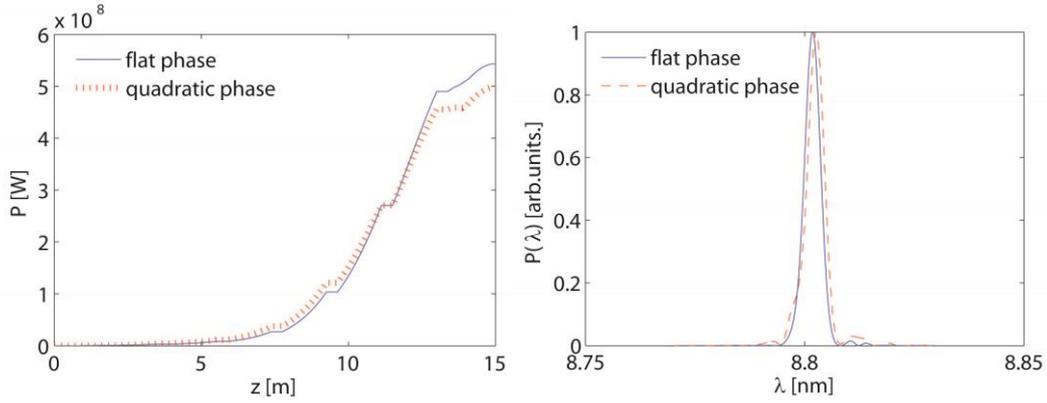

FIG.16. 30th harmonic radiation performances of EEHG with sub-harmonic modulator resonant at 1320 nm for the 80 fs seed laser pulses with flat phase (blue line) and quadratic phase (red dashed line): (a) FEL gain curves; (b) spectra at saturation.

# VI. CONCLUSIONS

In summary, energy modulations for seeded FEL schemes with chirped short seed laser pulses have been studied analytically and numerically. It is found that, by adopting a seed laser with pulse length comparable with the slippage length in the modulator, the initial phase error induced by the imperfect seed laser pulse can be significantly smoothed. 3D Simulations has been carried out and the results confirm the theoretical prediction. In this paper, only the slippage effects on linear frequency chirps in the seed pulse have been studied. We expect that the slippage effect can also be used to compensate other temporal and spatial errors induced by the seed laser with non-ideal properties, e.g. arbitrary spectral phase errors and laser field errors, etc. Further investigations on these issues will be performed in the future.

## ACKNOWLEDGEMENTS

The authors would like to thank J. Chen, T. Zhang, L. Shen and Z. Huang for helpful discussions and comments. This work is supported by Major State Basic Research Development Program of China (973 Program) (Grant No. 2011CB808300) and National Natural Science Foundation of China (Grant No. 10935011). The work of D. Xiang is supported by U.S. DOE Contract No. DE-AC02-76SF00515.